# Strong noise attenuation of seismic data based on Nash equilibrium

Mingwei Wang[1], Yingtian Liu[1], Junheng Peng[1], Yong Li, Huating Li[1]

1. School of Geophysics, Chengdu University of Technology, Chengdu, China

## Abstract

Seismic data acquisition is often affected by various types of noise, which degrade data quality and hinder subsequent interpretation. Recovery of seismic data becomes particularly challenging in the presence of strong noise, which significantly impacts both data accuracy and geological analysis. This study proposes a novel single-encoder, multiple-decoder network based on nash equalization (SEMD-Nash) for effective strong noise attenuation in seismic data. The main contributions of this method are as follows: First, we design a shared encoder-multi-decoder architecture, where an improved encoder extracts key features from the noisy data, and three parallel decoders reconstruct the denoised seismic signal from different perspectives. Second, we develop a multi-objective optimization system that integrates three loss functions—mean aquared error (MSE), perceived loss, and structural similarity index (SSIM)—to ensure effective signal reconstruction, high-order feature preservation, and structural integrity. Third, we introduce the nash equalization weight optimizer, which dynamically adjusts the weights of the loss functions, balancing the optimization objectives to improve the model's robustness and generalization. Experimental results demonstrate that the proposed method effectively suppresses strong noise while preserving the geological characteristics of the seismic data.

# 1. Introduction

Seismic exploration is one of the main geophysical methods in oil and gas resource exploration, and its data quality directly affects the accuracy of subsurface structure interpretation and the reliability of exploration decisions. However, in the actual exploration process, seismic data is often disturbed by noise, which will decrease the signal-to-noise ratio of the data, obscure valid information, and bring serious challenges to subsequent data processing and interpretation. Currently, the methods used to deal with noise can be broadly divided into two main categories: traditional computing and deep learning.

The traditional calculation methods are divided into three categories: spatial domain filtering, transformation domain filtering, and sparse representation filtering methods based on matrix downgrading. The spatial domain filtering method directly operates on the data in the spatial domain, and the filtering is achieved through convolution or other operations. The spatial domain method includes median filtering (Liu et al., 2009; Liu, 2013), mean filtering (Liu et al., 2006), etc. The transformation domain filtering method can convert seismic data into a sparse domain for better separation of signal and noise. Transform domain filtering methods include wavelet transform (Goudarzi et al., 2012), curved wave transform (Starck et al., 2002), shearlet transform (Hou et al., 2019), f-x prediction filter (Hornbostel, 1991), etc. The f-x prediction filter was originally used to attenuate random noise in post-stack seismic data, which assumes that linear seismic events are predictable in the frequency space (f-x) domain. The filtering method based on the sparsity representation of matrix downgrade is to perform sparse decomposition of seismic data, project the seismic signal in the time domain to the sparse domain, and obtain the sparsity coefficient, the larger the coefficient is the effective signal, the smaller coefficient is the noise, and the threshold value separates the effective signal and the noise. Filtering methods based on sparse representation based on matrix rank reduction include empirical mode decomposition (Flandrin et al., 2004) and singular value decomposition (Chen et al.,

2016).

In recent years, with the continuous advancement of computer technology, deep learning methods have been widely used in many fields (Hinton and Salakhutdinov, 2006). In geophysical noise attenuation, many scholars have proposed quite practical methods. Such as Saad et al. ( 2020) proposed a deep encoder noise attenuation method, in which time series seismic data is used as network input, and the network encodes the input seismic data into multiple levels of abstraction, These layers are then decoded to reconstruct a noise-free seismic signal. Peng et al. (2024b) proposed the use of diffusion models and principal component analysis to deal with seismic data noise. In addition, Peng et al. (2024a) also proposed a noise attenuation method for fast diffusion models, which achieves faster diffusion model denoising by improving Bayesian equations and new normalization methods. Yang et al. (2021b) used a deep-skip autoencoder to achieve unsupervised 3D random noise attenuation. In deep learning, convolutional neural networks (CNN), as the most commonly used basic networks for deep learning, have achieved remarkable results in the fields of computer vision and natural language processing (LeCun et al., 2015). Based on this, many researchers have proposed a scheme to use CNN for random noise attenuation. For example, Yang et al. (2021a) employed a CNN (ResNet) with residual connections for the suppression of random noise. Liao et al. (2023) designed a feedback structure based on ResNet, which gradually reduces the input noise level through multiple iterations, to extract the remaining signal and improve the denoising effect. Zhao et al. (2018) proposed the use of a denoising convolutional neural network (DnCNN) to directly learn noise extraction instead of attenuation. Peng et al. (2024) proposed an adaptive convolutional filter for seismic noise attenuation, which effectively attenuated the noise using three priors. Song et al. (2020) proposed the use of a deep convolutional autoencoder to suppress seismic noise, which was effectively suppressed by using multiple filters of different sizes in the encoding and decoding frames of the convolutional autoencoder. In general, many improvement methods based on deep learning have achieved certain results. However, when seismic data encounters strong noise, it is difficult to remove it.

Strong noise blurs the line between noise and signal, which can interfere with the encoder's accurate extraction of signal features. Conventional convolutional layers may not be effective at distinguishing between noise and signal. We are trying to find a concise and efficient network for strong noise attenuation of seismic data. We extract the features of the seismic data by building an improved convolutional encoder. But we also need to consider how to adjust the network using a suitable loss function. Therefore, we use multiple loss tasks to obtain a joint loss function that takes into account the characteristics of various aspects of maintaining seismic data. But we are faced with the question, how should we evaluate and adjust the weights of multiple losses? Therefore, we propose to use Nash equilibrium to adjust the weights between the losses of multiple tasks. We will explain the specifics in the methodology section of the article.

# 2. Method

Our network consists of an encoder, three parallel decoders that read shared features, and a Nash weight optimizer. We will explain our approach from these three aspects.

## 2.1 Encoder

We use the convolution layer + Batch Normalization layer + ReLU activation function for the encoder. As shown in Figure 1, we are in the second and third convolutional layers, and we replace the convolutional layers with deformable convolutional layers. As shown in Figure 2, compared with the traditional convolutional layer, the deformable convolutional layer has an additional offset, which can add a new receptive field in addition to the receptive field extracted by the traditional convolution. Therefore, deformable convolutions are more sensitive to the features of the data and can better capture the complex structural features of the target (Dai et al., 2017; Mingwei et al., 2024). For any point $p_0$ on the input feature map, the ordinary convolution can be expressed as

$$y(p_0) = \sum_{p_n \in R} w(p_n) * x(p_0 + p_n)$$

where $p_0$ represents the offset of each point in the convolutional kernel relative to the center point. The $w(p_n)$ resents the weight of the corresponding position of the convolution kernel. The $x(p_0 + p_n)$ resents the value of the element at the position on the input feature plot $p_0 + p_n$. The $y(p_0)$ presents the value of the element at the location on the output feature plot $p_0$, it is obtained by convolution kernel and input feature map. Deformable convolution introduces an offset $\Delta p_n$ compared to traditional convolution, deformable convolution can be expressed as

$$y(p_0) = \sum_{p_n \in R} w(p_n) * x(p_0 + p_n + \Delta p_n)$$

Therefore, deformable convolution has stronger feature extraction capabilities than traditional convolution, and we apply it to distinguish noise from effective signals in the case of strong noise.

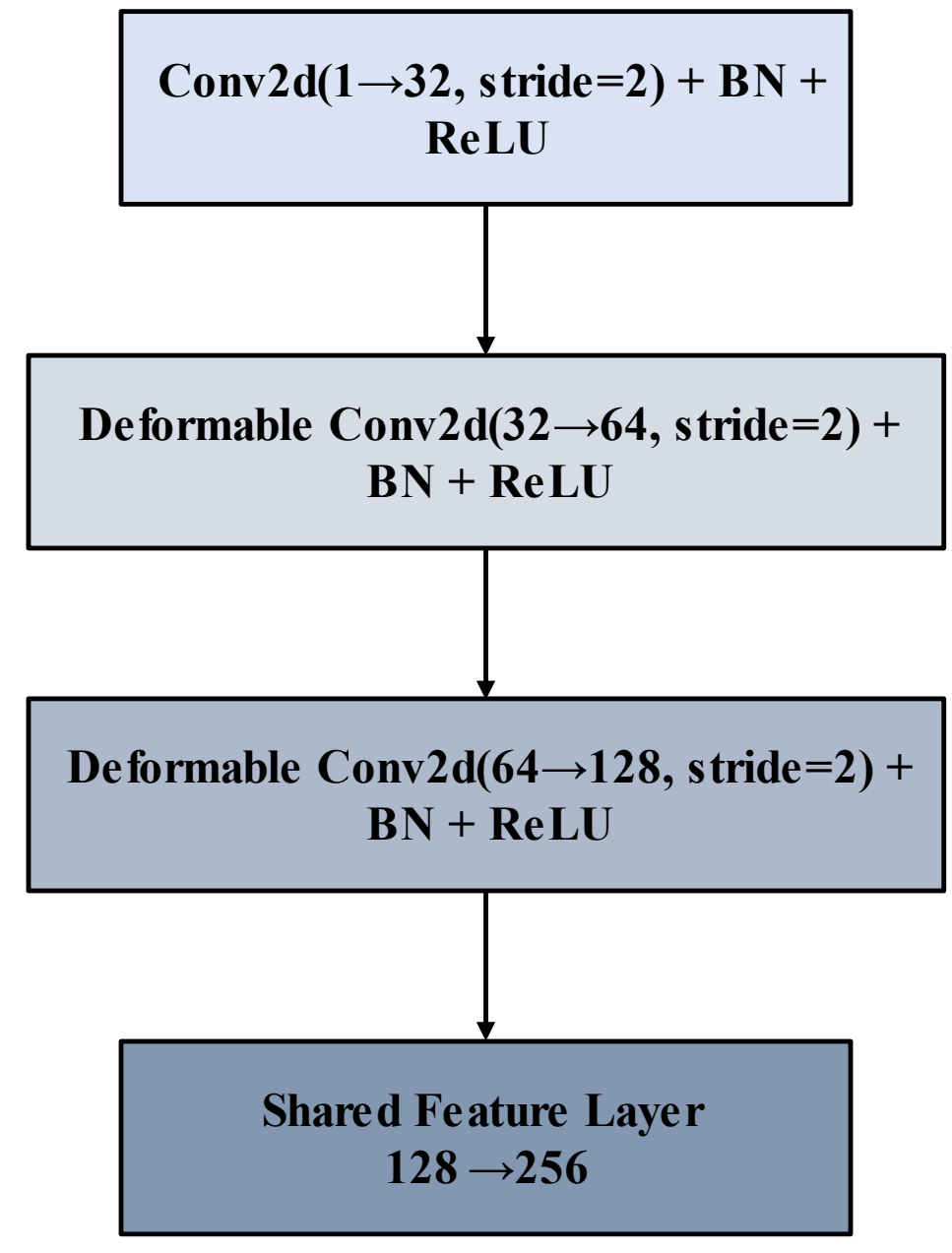

**Figure 1.** Structure of encoder.

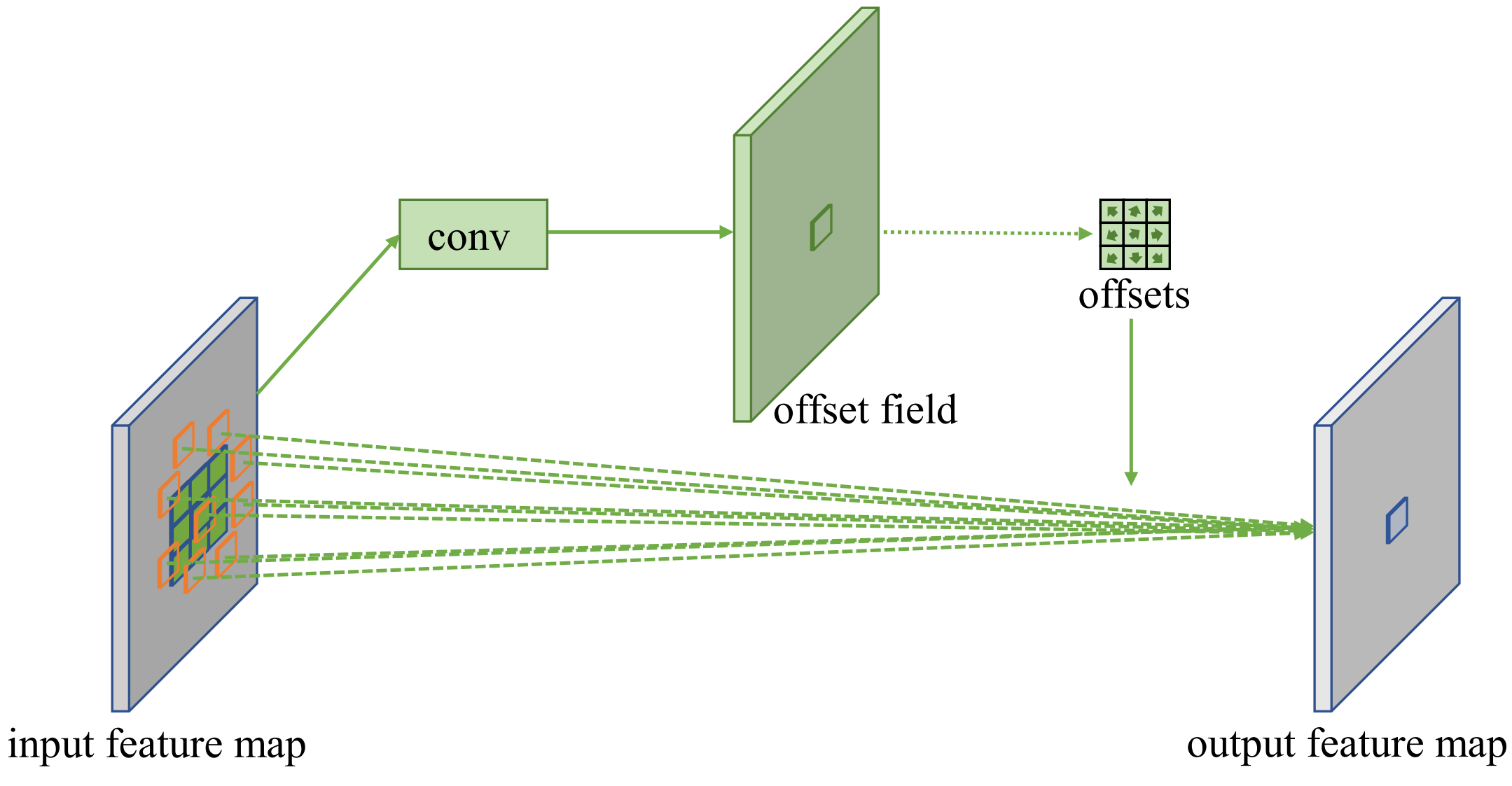

**Figure 2.** Structure of deformable convolution.

### 2.2 Decoder

In the decoder part, we use a multi-task structure with three decoders in parallel. The three decoders share the features extracted by the encoder but use different loss functions. So, the final loss function can combine the advantages of the three loss functions. The network structure is shown in Figure 3, and the features transmitted by the encoder are shared by three decoders, which use the MSE loss function (Chicco et al., 2021), the Perceptual loss function (Yang et al., 2018) and the SSIM loss function. The MSE loss function is responsible for ensuring the accuracy of the overall data reconstruction and reducing the impact of noise. The Perceptual loss function is responsible for maintaining the high-level characteristics of seismic data and preserving the key features of seismic data construction. The SSIM loss function is responsible for maintaining the structural information of the seismic data and increasing the continuity of the seismic data horizon. The three losses were finally merged into a joint loss by Nash weight optimization. where the weight of the joint loss satisfies

$$Loss\ weight = \lambda_1(MSE\ Loss) + \lambda_2(Perceptual\ Loss) + \lambda_3(SSIM\ Loss) = 1$$

By using joint loss, we combine the advantages of these three loss functions, so that the network can still preserve all aspects of seismic data in the face of strong noise.

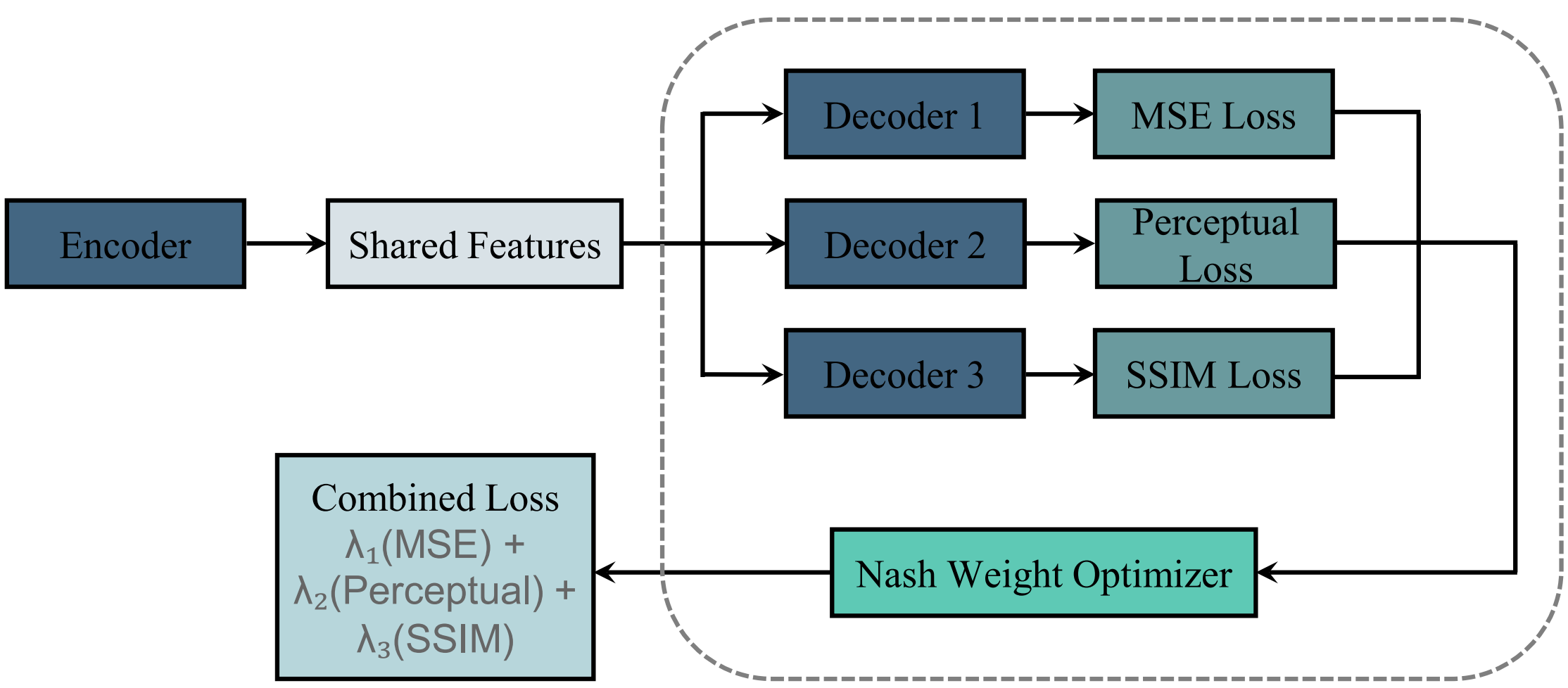

**Figure 3.** Structure of decoder.

### 2.3 Nash Weight Optimizer

Nash equilibrium is an important concept in game theory, which is widely used in many fields such as economics, computer science, and artificial intelligence (Daskalakis et al., 2009; Liu et al., 2024). It describes how to choose an optimal strategy in a game taking into account other tasks. In Nash equilibrium, there is a utility function and an optimal response function. Utility functions are used to describe the "benefit" that each task will receive when choosing a strategy. Utility functions help analyze how individual tasks make decisions based on their preferences and goals to maximize their benefits or utility. The best response function refers to the optimal strategy chosen by a task to maximize its utility (or benefit) given other task strategies. The best response function represents the optimal choice of another task when the strategy of other tasks is fixed.

In this network, we regard the weights of the three parallel loss functions as three parallel tasks that need to be optimized and use Nash to find an optimal strategy. We define the MSE loss function as $L_m$, the perceived loss function as $L_p$, the SSIM loss function as $L_s$, the MSE loss function weight as $w_m$, the perceived loss function weight as $w_p$, and the SSIM loss function weight as $w_s$. The utility function is defined as

$$U_m(w_m, w_p, w_s) = -\left(w_m * L_m + \gamma * \left(w_p * L_p + w_s * L_s\right)\right) \quad (1)$$

$$U_p(w_m, w_p, w_s) = -\left(w_p * L_p + \gamma * (w_m * L_m + w_s * L_s)\right) \quad (2)$$

$$U_s(w_m, w_p, w_s) = -\left(w_s * L_s + \gamma * \left(w_m * L_m + w_p * L_p\right)\right) \quad (3)$$

where $\gamma$ is the interaction coefficient, which controls the degree of influence between tasks. $U_m$ denotes the utility function of the MSE loss weight, $U_p$ denotes the utility function of the perceived loss weight, $U_s$ denotes the utility function of the SSIM loss weight. The Nash equilibrium conditions we are looking for are

$$\partial U_m / \partial w_m = 0 \quad (4)$$

$$\partial U_p / \partial w_p = 0 \quad (5)$$

$$\partial U_s / \partial w_s = 0 \quad (6)$$

Bringing equations 1, 2, 3 into equations 4, 5, 6, there are

$$-L_m - \gamma * \left(w_p * \partial L_p / \partial w_m + w_s * \partial L_S / \partial w_m\right) = 0 \quad (7)$$

$$-L_p - \gamma * \left(w_m * \partial L_m / \partial w_p + w_s * \partial L_S / \partial w_p\right) = 0 \quad (8)$$

$$-L_S - \gamma * \left(w_m * \partial L_m / \partial w_s + w_p * \partial L_p / \partial w_s\right) = 0 \quad (9)$$

Where there is a constraint because the sum of the weights of the loss functions is 1 and all of them are greater than or equal to 0:

$$w_m + w_p + w_s = 1$$

$$w_m, w_p, w_s \geq 0$$

The weight update formula for the loss function is

$$w_m^{(t+1)} = w_m^t + \eta * (\partial U_m / \partial w_m) \tag{10}$$

$$w_p^{(t+1)} = w_p^t + \eta * \left(\partial U_p / \partial w_p\right) \tag{11}$$

$$w_s^{(t+1)} = w_s^t + \eta * (\partial U_s / \partial w_s) \tag{12}$$

where t refers to the t-th weight update, and t+1 is the (t+1)-th weight update, $\eta$ indicates the learning rate. The best response function is expressed as:

$$BR_m\left(w_p, w_s\right) = argmax_{w_m}\, U_m\left(w_m, w_p, w_s\right) \tag{13}$$

$$BR_p(w_m, w_s) = argmax_{w_p}\, U_p\left(w_m, w_p, w_s\right) \tag{14}$$

$$BR_s\left(w_m, w_p\right) = argmax_{w_s}\, U_s\left(w_m, w_p, w_s\right) \tag{15}$$

In equation 13, 14 and 15, $BR_m$, $BR_p$, and $BR_s$ represent the best response function for the weight of the MSE loss function, the best response function for the weight of the perceived los function, and the best response function for the weight of the SSIM loss function, respectively. Equation 13 represents the selection of an optimal weight $w_m$ for the MSE loss function so that its utility $U_m\left(w_m, w_p, w_s\right)$ is maximum, and equation 14 and 15 are similar to equation 13. We denote the optimal weight $w_m$ as $w^*{}_m$, the optimal weight $w_p$ as $w^*{}_p$, and the optimal weight $w_s$ as $w^*{}_s$. Nash equilibrium satisfies

$$w^*{}_m = BR_m\left(w^*{}_p, w^*{}_s\right) \tag{16}$$

$$w^*{}_p = BR_p(w^*{}_m, w^*{}_s) \tag{17}$$

$$w^*{}_s = BR_s\left(w^*{}_m, w^*{}_p\right) \tag{18}$$

Finally, we determine the conditions under which the Nash equilibrium is optimized, i.e., the conditions that the values $w^*{}_m$, $w^*{}_p$ and $w^*{}_s$ must satisfy are

$$\left|w_m^{(t+1)} - w_m^t\right| < \varepsilon \tag{19}$$

$$\left|w_p^{(t+1)} - w_p^t\right| < \varepsilon \tag{20}$$

$$\left|w_s^{(t+1)} - w_s^t\right| < \varepsilon \tag{21}$$

Here, $\varepsilon$ represents the minimum convergence threshold, indicating that the weights of our loss functions have converged to a stable state. Specifically, when the absolute difference between the weights of the network at the t-th and (t+1)-th iterations is smaller than $\varepsilon$, we determine that the Nash network has reached a convergent state.

In summary, we designed an improved encoder and a parallel decoder structure with multi-task loss functions (SEMD-Nash). By using Nash equilibrium, we obtained the optimal values for the weights between the three loss functions, resulting in an optimal weighted mixed loss function. This led to favorable results, which will be presented in detail in the experimental section.

# 3 Experiment

## 3.1 Experimental equipment and parameters

Our lab equipment is an NVIDIA RTX 3060 Ti GPU with 8GB memory. The network parameters of SEMD-Nash are shown in Table 1.

**Table 1.** Network parameters.

| Parameter | Value |
|---|---|
| Epoch | 200 |
| Batch size | 16 |
| Learning rate($\eta$) | 0.001 |
| Weight update frequency | 1 |
| Minimum convergence ($\varepsilon$) | 0.0001 |

For comparison methods, our first approach selects the well-established f-x deconvolution (f-x decon) (Canales, 1984), the second approach adopts the Local Damped Rank Reduction (LDRR) by Chen et al. (2023), and the third method utilizes the Twice Denoising Autoencoder Framework for Random Seismic Noise Attenuation (TDAE) by Liao et al (2023).

## 3.2 Synthetic examples

As shown in Figure 4a, the first synthetic example contains a continuous geological structure, which helps us to observe leakage during random noise attenuation. Figure 4b shows the noise input, and we add a strong noise of 1.5 standard deviations to the clean data, resulting in a signal-to-noise ratio (SNR) of -2.525 dB for the noise input. The SNR formula can be expressed as:

$$SNR = 10 * \log_{10} \frac{||x_L||^2}{||x_N - x_L)||^2}$$

where $x_L$ represents clean seismic data, and $x_N$ represents noisy seismic data. The SNR is an effective measure of noise attenuation performance and signal retention for various methods. As shown in Figure 4, SEMD-Nash (Figure 4f) is the cleanest and removes the most noise. Compared to SEMD-Nash (Figure 4f), there is still partial random noise in f-x decon (Figure 4c), LDRR (Figure 4d), and TDAE (Figure 4e). As can also be seen from Table 2, SEMD-Nash has the highest SNR.

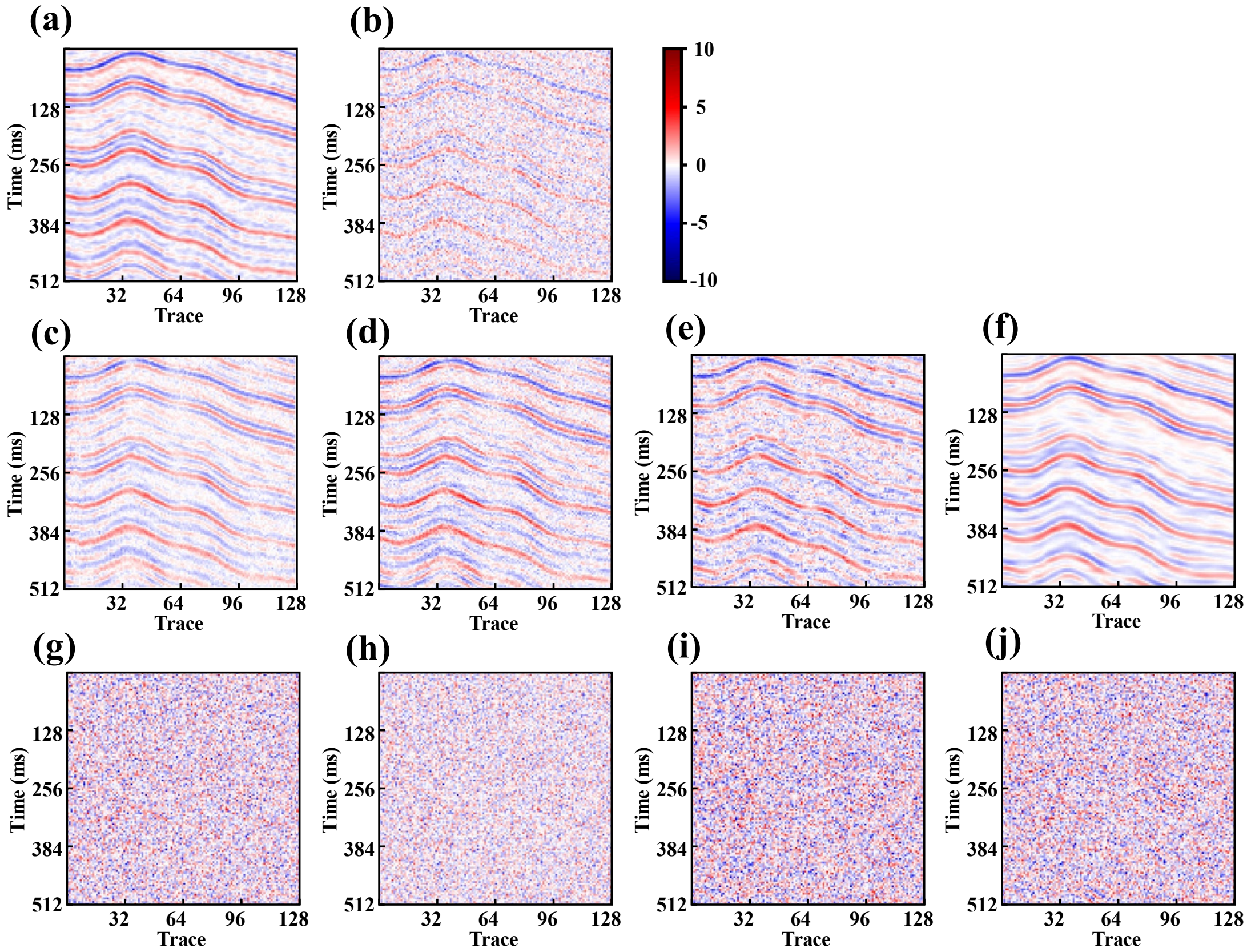

**Figure 4.** Synthetic example 1 and comparisons among f-x decon, LDRR, TDAE, and SEMD-Nash. (a) Synthetic Example 1. (b) Noisy data. (c-f) Estimated outputs of f-x decon, LDRR, TDAE, and SEMD-Nash, respectively. (g-j) Estimated noise distributions of f-x decon, LDRR, TDAE, and SEMD-Nash, respectively.

**Table 2.** SNR of synthetic examples. Bold values represent the best SNR performance of each example.

| | Noise data | f-x decon | LDRR | TDAE | SEMD-Nash |
|---|---|---|---|---|---|
| Synthetic example 1 | -2.525 | 6.114 | 5.295 | 4.204 | **7.010** |
| Synthetic example 2 | -5.346 | 5.169 | 2.976 | -1.699 | **6.672** |

In the process of random noise attenuation, it is also possible to cause the loss of effective signal. As can be seen in Figure 4c-f, there is signal leakage for each method. To observe the loss of effective signals more effectively, we introduced local similarity (Chen and Fomel, 2015) to evaluate the signal preservation of various methods in the work. Local similarity is widely used in various seismic data noise processing tasks to evaluate the similarities between attenuated noise and processed data (signal leakage). As shown in Figure 5, the local similarity was the best for SEMD-Nash (Figure 5d), the worst for LDRR (Figure 5b), followed by TDAE (Figure 5c), and then f-x decon (Figure 5a). Table 3 shows the local similarity values of the synthetic data, which also correspond to the local similarities in the figure.

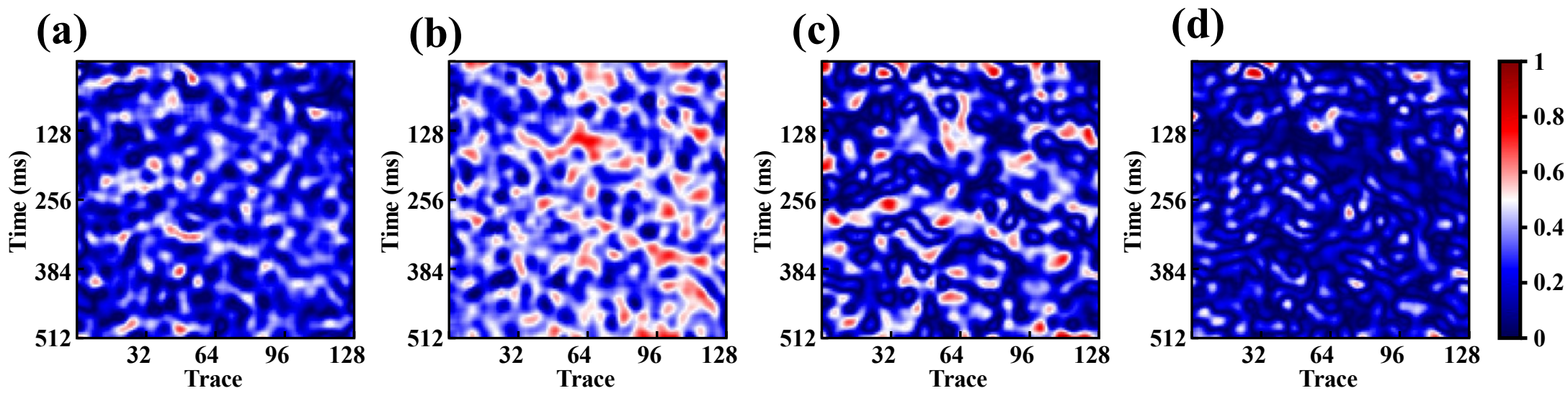

**Figure 5.** Local similarity in synthetic example 1. (a-d) The local similarities of f-x decon, LDRR, TDAE, and SEMD-Nash, respectively.

**Table 3.** Local similarity of synthetic examples. Bold values represent the best local similarity performance of each example.

| | f-x decon | LDRR | TDAE | SEMD-Nash |
|---|---|---|---|---|
| Synthetic example 1 | 0.251 | 0.374 | 0.260 | **0.163** |
| Synthetic example 2 | 0.286 | 0.433 | 0.416 | **0.154** |

As shown in Figure 6a, there are some faults in synthetic example 2, which can test how the network behaves on discontinuous data. Figure 6b is the noise input, and we add a strong random noise with a standard deviation of 1.8 relative to itself to synthesis example 2, resulting in an SNR of -5.346 dB, which is well above the level of strong random noise. As can be seen in Figure 6b, the data is affected by strong random noise, and most of the effective signal is already covered by noise, which is a challenge for the various methods. Compared to SEMD-Nash (Figure 6f), the f-x decon (Figure 6c), LDRR (Figure 6d), and TDAE (Figure 6e) are still partially noisy, resulting in the active signal being covered by noise. However, compared to the clean data (Fig. 6a), there is still signal leakage in SEMD-Nash (Fig. 6f), and some of the curve structures are not smooth due to the absence of valid signal.

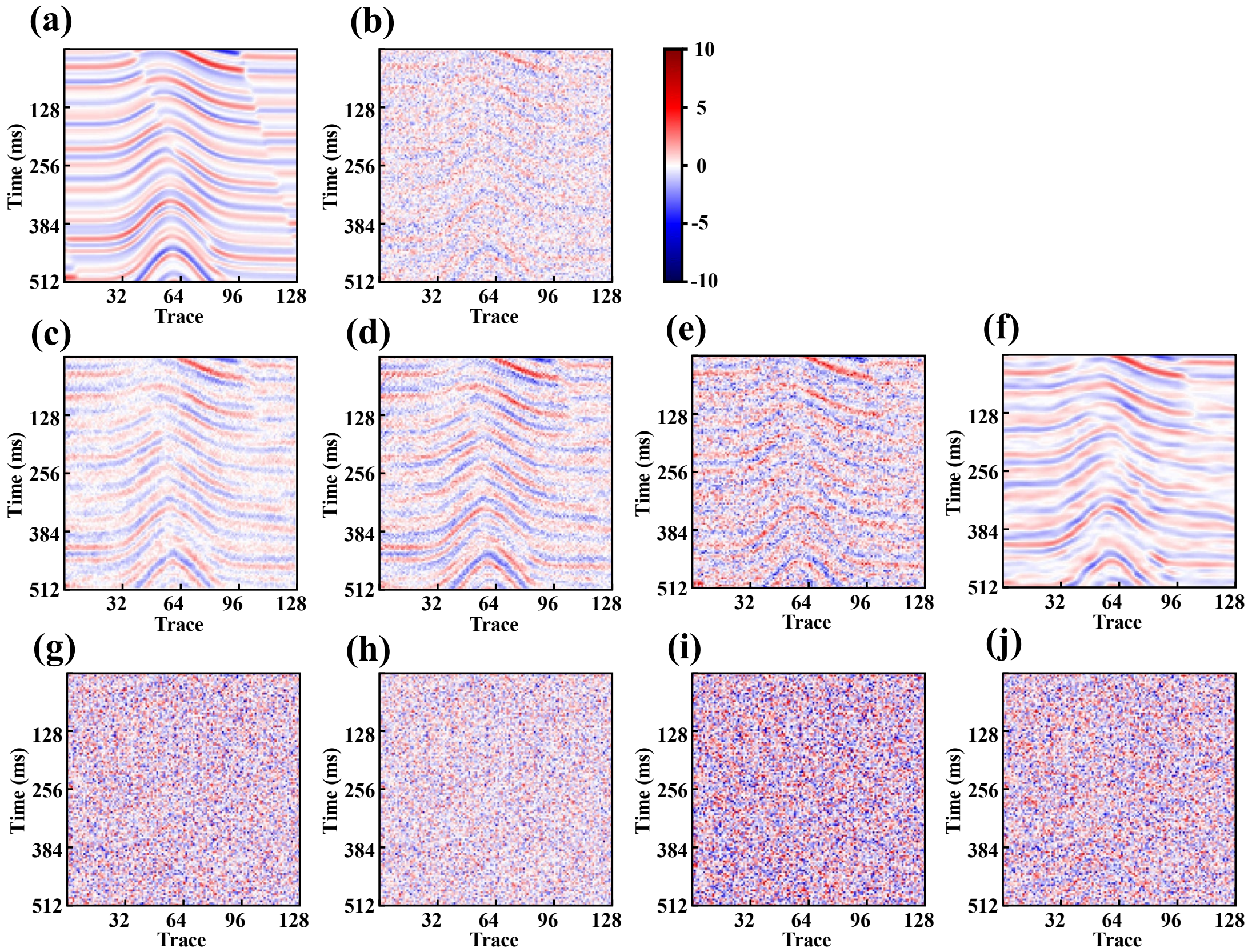

**Figure 6.** Synthetic example 2 and comparisons among f-x decon, LDRR, TDAE, and SEMD-Nash. (a) synthetic example 1. (b) noisy data. (c-f) estimated outputs of f-x decon, LDRR, TDAE, and SEMD-Nash, respectively. (g-j) estimated noise distributions of f-x decon, LDRR, TDAE, and SEMD-Nash, respectively.

To visualize the signal leakage of each method, we plotted a local similarity plot for Synthesis Example 2, as shown in Figure 7. LDRR and TDAE have a large number of effective signal leaks, and f-x decon and SEMD-Nash have less signal leakage, but SEMD-Nash has the least signal leakage.

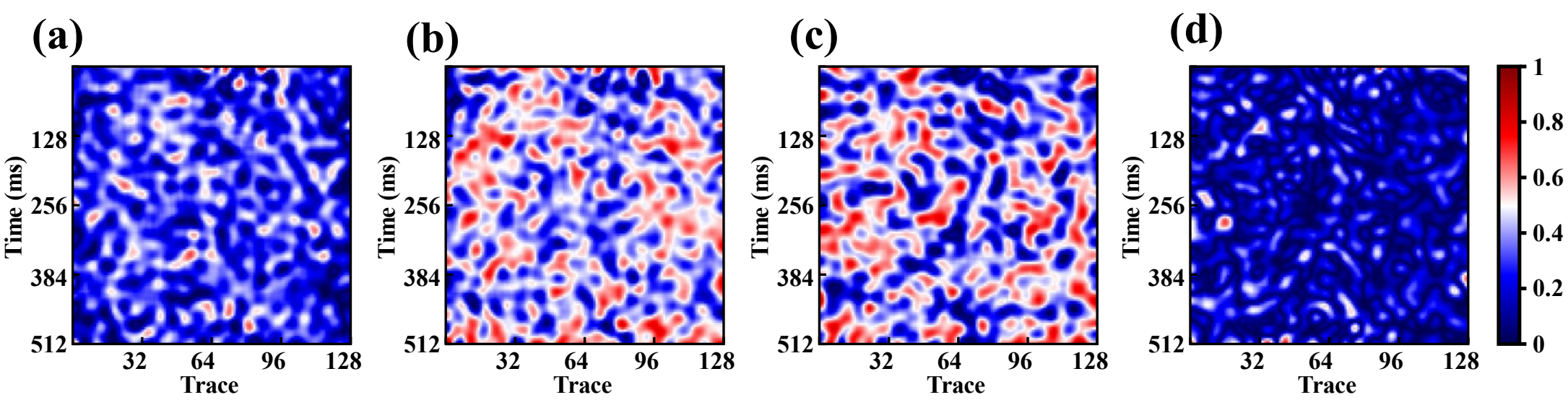

**Figure 7.** Local similarity in synthetic example 2. (a-d) the local similarities of f-x decon, LDRR, TDAE, and SEMD-Nash, respectively.

From Tables 2 and 3, SEMD-Nash has the highest SNR and better local similarity in the synthetic example compared to f-x decon, LDRR, and TDAE. This shows that there is only a small amount of effective signal leakage in SEMD-Nash when a large amount of noise is removed. In addition, four methods were used for more synthesis examples. We processed 50 identical synthesis examples each using f-x decon, LDRR, TDAE, and SEMD-Nash. As shown in Figure 8, we plotted a violin plot concerning the SNR. In the violin diagram, the median SNR of SEMD-Nash is the highest, close to 14 dB, which indicates that the SNR of SEMD-Nash has a higher concentrated trend. It can also be seen from the average SNR that SEMD-Nash has the highest average SNR (13.22dB).

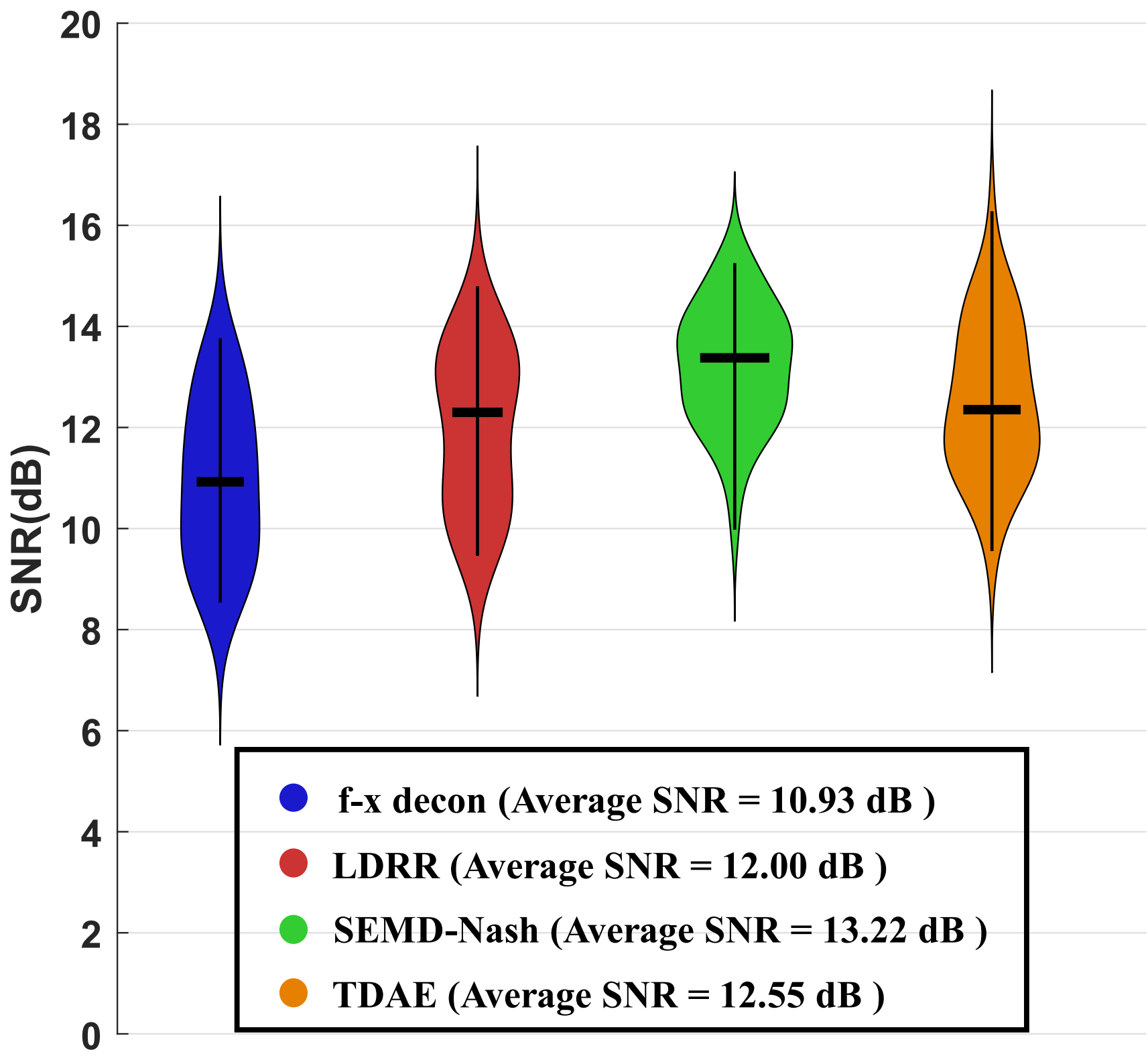

**Figure 8.** The SNR of the synthetic examples.

### 3.3 Field examples

In the field of geophysics, field data acquisition is fundamental to understanding geophysical characteristics. However, unlike synthesized clean data, field data is often noisy. Therefore, how to achieve noise attenuation and extract effective signals from field data is an important problem in seismic data processing. We selected field data from a region as our field example, both of which included 128-time samples at a distance of 4160 m.

As shown in Figure 9a, field example 1 is covered by a large amount of noise, which makes the extraction of valid signals difficult. We used the f-x decon, LDRR, TDAE, and SEMD-Nash methods for field examples. TDAE (Figure 9d) did not perform well on field example 1, with only a small amount of noise removed (Figure 9h). As you can see from the noise graph, there is some signal leakage in the f-x decon (Figure 9f) and LDRR (Figure 9g). Although SEMD-Nash (Figure 9i) also has signal leakage, it achieves better results compared to f-x decon and LDRR.

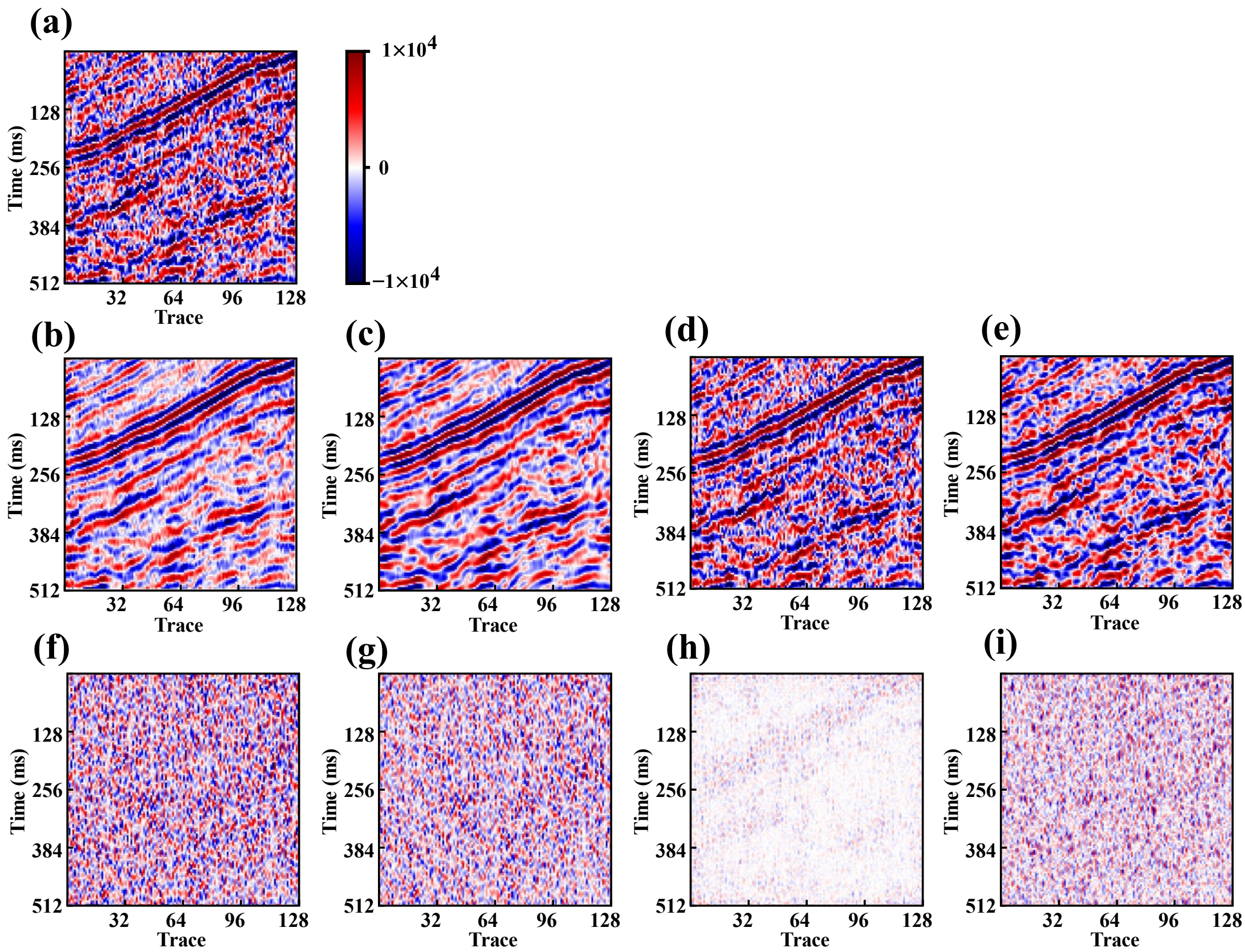

**Figure 9.** Field example 1. (a) field example. (b-e) estimated outputs of f-x decon, LDRR, TDAE, and SEMD-Nash, respectively. (f-i) estimated noise distributions of f-x decon, LDRR, TDAE, and SEMD-Nash, respectively.

To better evaluate the methods, we also used local similarities. As shown in Figure 10 and Table 4, SEMD-Nash has the best local similarity, followed by LDRR, f-x decon, and TDAE.

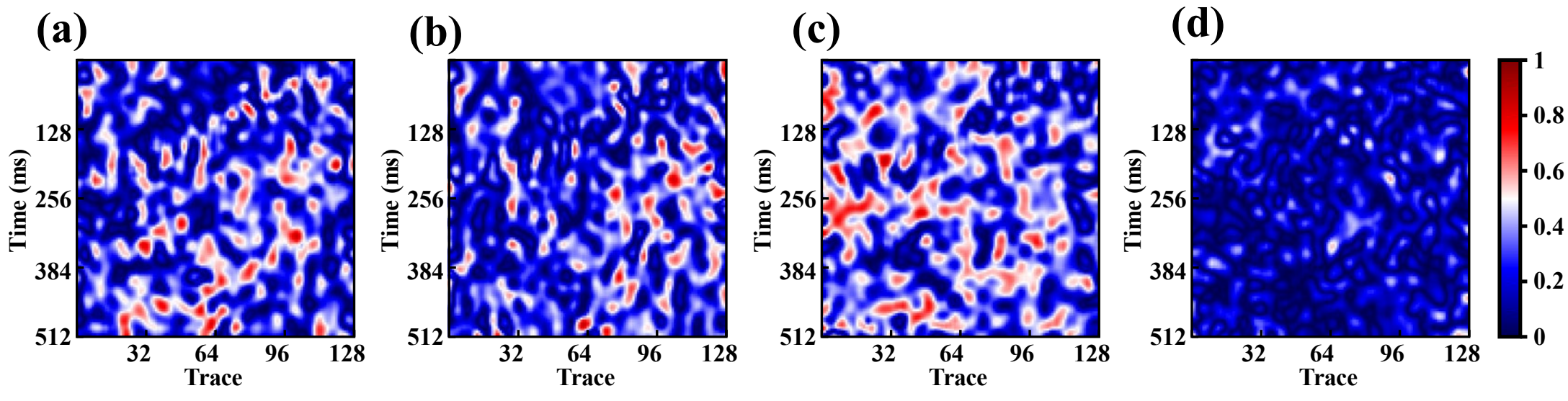

**Figure 10.** Local similarity in field example 1. (a-d) the local similarities of f-x decon, LDRR, TDAE, and SEMD-Nash, respectively.

**Table 4.** Local similarity of field examples. Bold values represent the best local similarity performance of each example.

| | f-x decon | LDRR | TDAE | SEMD-Nash |
|---|---|---|---|---|
| field example1 | 0.288 | 0.272 | 0.342 | **0.144** |
| field example2 | 0.359 | 0.327 | 0.467 | **0.153** |

As shown in Figure 11a, field example 2 has some faulty structures but is covered by noise, which makes it more difficult for the network to distinguish between noise and valid signals. As can be seen from Figure 11b-e, f-x decon (Figure 11b), LDRR (Figure 11c), and SEMD-Nash (Figure 11e) remove most of the random noise, and TDAE (Figure 11d) removes a small amount of random noise. As can be seen from the noise plot (Figure 11f-i), there is signal leakage between channels 64 and 96 and channels 96 and 128 for each method. Among them, SEMD-Nash (Fig. 11i) removes a large amount of random noise while reducing signal leakage.

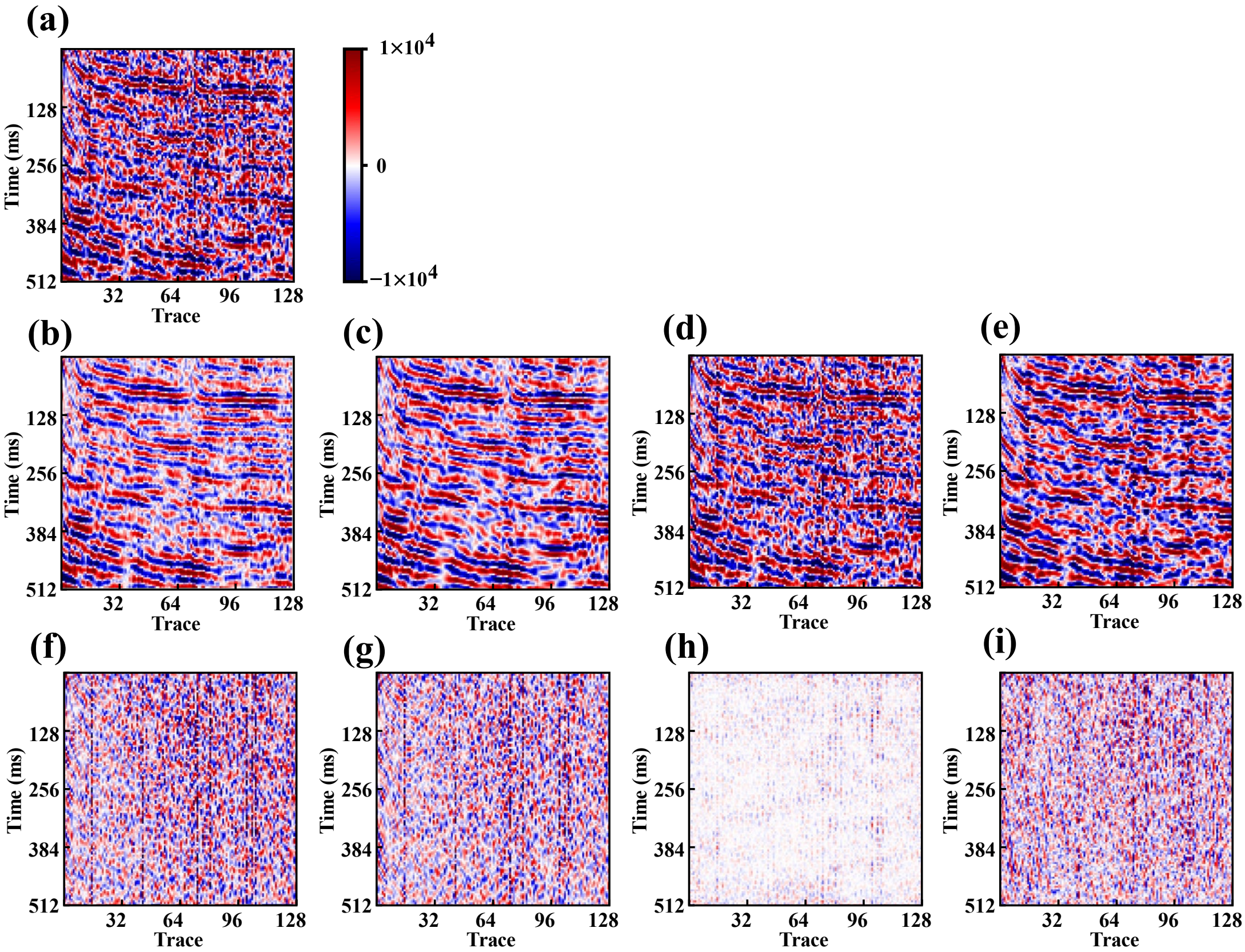

**Figure 11.** Field example 2. (a) field example. (b-e) estimated outputs of f-x decon, LDRR, TDAE, and SEMD-Nash, respectively. (f-i) estimated noise distributions of f-x decon, LDRR, TDAE, and SEMD-Nash, respectively.

We used local similarity as an evaluation metric, as shown in Figure 11 and Table

4, TDAE had the worst local similarity, followed by LDRR, and then f-x decon. The local similarity of SEMD-Nash is the best. Combining Figures 11 and 12, SEMD-Nash retains a much better valid signal while removing random noise than f-x decon, LDRR, and TDAE.

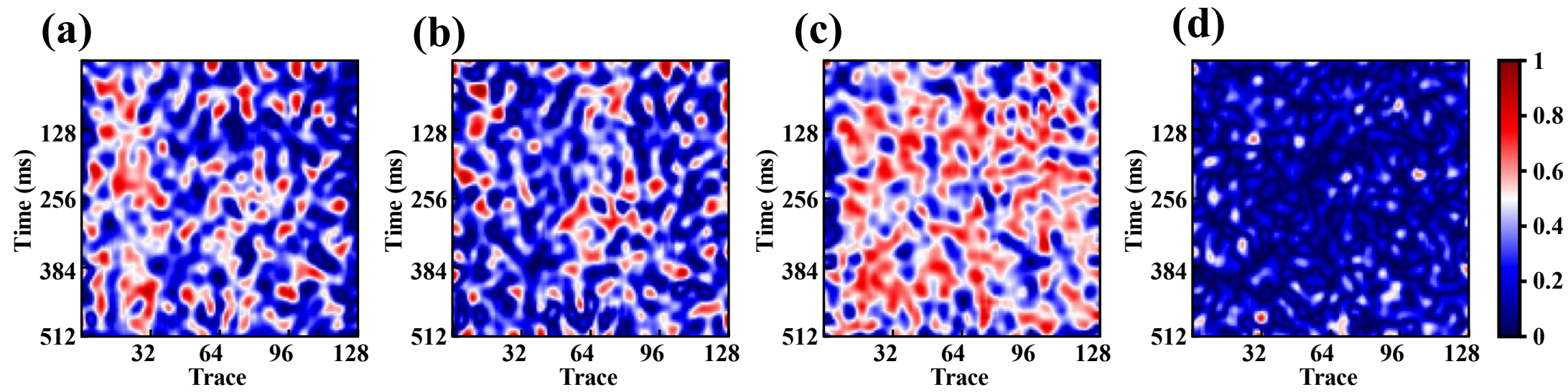

**Figure 12.** Local similarity in field example 1. (a-d) the local similarities of f-x decon, LDRR, TDAE, and SEMD-Nash, respectively.

# 4 Discussion

In the model with multiple loss functions, the weight initialization has a great impact on the training process of the network. The weights of each loss function control the degree to which the model pays attention to different objectives during the optimization process. Reasonable initialization methods can help the model converge better, while improper initialization may lead to too much or too little influence of some loss functions, which in turn affects the performance of the final model. While SEMD-Nash combines the advantages of the three loss functions, it also faces the problem of initialization of the weights of the three loss functions. Proper initialization loss function weights can make the Nash weight optimizer converge faster, allowing it to converge with fewer epochs of training. The numerical range of one loss function is much larger than that of other loss functions, and if left unadjusted, it may cause the loss function to dominate the training process, which in turn will cause the training to be biased in one aspect and affect the global performance of the model. Therefore, we initialize the weights by averaging the weights. This is shown in Table 5. We tested a bias towards a loss function as the dominant training network, which would increase the epoch of the Nash weight optimizer convergence. We haven't found a better way to assign the initial weights of the loss function at the moment. In the end, we chose to initialize the loss function weights using the average weights. In the following research, we will continue to look for a better way to establish the initial weights of the loss function.

**Table 4.** Initial weight setting of the loss function.

| $\lambda_1$ | $\lambda_2$ | $\lambda_3$ | **Epoch** |
|---|---|---|---|
| 1 | 0 | 0 | **94** |
| 0 | 1 | 0 | **113** |
| 0 | 0 | 1 | **142** |
| 1/3 | 1/3 | 1/3 | **67** |

## 5 Conclusion

The single-encoder multi-decoder network based on Nash equalization (SEMD-Nash) proposed in this paper provides an effective solution for the strong noise attenuation of seismic data. By designing an improved network architecture with encoders and multiple decoders in parallel, SEMD-Nash can reconstruct clear seismic signals from multiple angles and effectively extract the essential characteristics of the noise data. In addition, the multi-objective optimization system constructed in this paper integrates three loss functions, namely Mean Square Error (MSE), Perceived Loss, and Structural Similarity (SSIM), to ensure the reconstruction of signal details, the maintenance of higher-order features, and the restoration of structural integrity. More importantly, by introducing the Nash equilibrium weight optimizer, the proposed method can dynamically adjust the weights of different loss functions during the training process, to achieve adaptive equilibrium for multi-objective optimization. Experimental results show that the proposed method can effectively suppress strong noise interference, and improve the accuracy and quality of seismic data recovery based on maintaining the original characteristics of seismic data.

# References

Anvari, R., Kahoo, A. R., Monfared, M. S., Mohammadi, M., Omer, R. M. D., & Mohammed, A. H. (2021). Random noise attenuation in seismic data using Hankel sparse low-rank approximation. *Computers & Geosciences*, *153*, 104802.

Canales, L. L. (1984). Random noise reduction. In *SEG Technical Program Expanded Abstracts 1984* (pp. 525-527). Society of Exploration Geophysicists.

Chen, Y., & Fomel, S. (2015). Random noise attenuation using local signal-and-noise orthogonalization. *Geophysics*, *80*(6), WD1-WD9.

Chen, Y., Huang, W., Yang, L., Oboué, Y. A. S. I., Saad, O. M., & Chen, Y. (2023). DRR: An open-source multi-platform package for the damped rank-reduction method and its applications in seismology. *Computers & Geosciences*, *180*, 105440.

Chen, Y., Zhang, D., Jin, Z., Chen, X., Zu, S., Huang, W., & Gan, S. (2016). Simultaneous denoising and reconstruction of 5-D seismic data via damped rank-reduction method. *Geophysical Journal International*, *206*(3), 1695-1717.

Chicco, D., Warrens, M. J., & Jurman, G. (2021). The coefficient of determination R-squared is more informative than SMAPE, MAE, MAPE, MSE and RMSE in regression analysis evaluation. *Peerj computer science*, *7*, e623.

Dai, J., Qi, H., Xiong, Y., Li, Y., Zhang, G., Hu, H., & Wei, Y. (2017). Deformable convolutional networks. In *Proceedings of the IEEE international conference on computer vision* (pp. Daskalakis, C., Goldberg, P. W., & Papadimitriou, C. H. (2009). The complexity of computing a Nash equilibrium. *Communications of the ACM*, *52*(2), 89-97.

Flandrin, P., Goncalves, P., & Rilling, G. (2004, September). Detrending and denoising with empirical mode decompositions. In *2004 12th European signal processing conference* (pp. 1581-1584). IEEE.

Goudarzi, A., & Riahi, M. A. (2012). Seismic coherent and random noise attenuation using the undecimated discrete wavelet transform method with WDGA technique. *Journal of Geophysics and Engineering*, *9*(6), 619-631.

Hinton, G. E., & Salakhutdinov, R. R. (2006). Reducing the dimensionality of data with neural networks. *science*, *313*(5786), 504-507.

Hornbostel, S. (1991). Spatial prediction filtering in the tx and fx domains. *Geophysics*, *56*(12), 2019-2026.

Hou, W. L., Jia, R. S., Sun, H. M., Zhang, X. L., Deng, M. D., & Tian, Y. (2019). Random noise reduction in seismic data by using bidimensional empirical mode decomposition and shearlet transform. *IEEE Access*, *7*, 71374-71386.

LeCun, Y., Bengio, Y., & Hinton, G. (2015). Deep learning. *nature*, *521*(7553), 436-444.

Liao, Z., Li, Y., Xia, E., Liu, Y., & Hu, R. (2023). A twice denoising autoencoder framework for random seismic noise attenuation. *IEEE Transactions on Geoscience and Remote Sensing*, *61*, 1-15.

Liu, C., Liu, Y., Yang, B., Wang, D., & Sun, J. (2006). A 2D multistage median filter to reduce random seismic noise. *Geophysics*, *71*(5), V105-V110.

Liu, Y. (2013). Noise reduction by vector median filtering. *Geophysics*, *78*(3), V79-V87.

Liu, Y., Li, Y., Li, H., Peng, J., Liao, Z., & Feng, W. (2024). The Nash-MTL-STCN for Prestack Three-Parameter Inversion. *arXiv preprint arXiv:2407.00684*.

Liu, Y., Liu, C., & Wang, D. (2009). A 1D time-varying median filter for seismic random, spike-like noise elimination. Geophysics, 74(1), V17-V24.

Mingwei, W., Yong, L., Yingtian, L., Junheng, P., & Huating, L. (2024). DCMSA: Multi-Head Self-Attention Mechanism Based on Deformable Convolution For Seismic Data Denoising. *arXiv preprint arXiv:2408.06963*.

Peng, J., Li, Y., Liao, Z., Wang, X., & Yang, X. (2024). Seismic Data Strong Noise Attenuation Based on Diffusion Model and Principal Component Analysis. *IEEE Transactions on Geoscience and Remote Sensing*.

Peng, J., Li, Y., Liu, Y., & Liao, Z. (2024). Fast Diffusion Model For Seismic Data Noise Attenuation. *arXiv preprint arXiv:2404.02767*.

Peng, J., Li, Y., LIu, Y., Wang, M., & Li, H. (2024). Adaptive Convolutional Filter for Seismic Noise Attenuation. *arXiv preprint arXiv:2410.18896*.

Saad, O. M., & Chen, Y. (2020). Deep denoising autoencoder for seismic random noise attenuation. *Geophysics*, *85*(4), V367-V376.

Song, H., Gao, Y., Chen, W., Xue, Y. J., Zhang, H., & Zhang, X. (2020). Seismic random noise suppression using deep convolutional autoencoder neural network. *Journal of Applied Geophysics*, *178*, 104071.

Starck, J. L., Candès, E. J., & Donoho, D. L. (2002). The curvelet transform for image denoising. *IEEE Transactions on image processing*, *11*(6), 670-684.

Yang, L., Chen, W., Wang, H., & Chen, Y. (2021). Deep learning seismic random noise attenuation via improved residual convolutional neural network. *IEEE Transactions on Geoscience and Remote Sensing*, *59*(9), 7968-7981.

Yang, L., Wang, S., Chen, X., Saad, O. M., Chen, W., Oboue, Y. A. S. I., & Chen, Y. (2021). Unsupervised 3-D random noise attenuation using deep skip autoencoder. *IEEE Transactions on Geoscience and Remote Sensing*, *60*, 1-16.

Yang, Q., Yan, P., Zhang, Y., Yu, H., Shi, Y., Mou, X., ... & Wang, G. (2018). Low-dose CT image denoising using a generative adversarial network with Wasserstein distance and perceptual loss. *IEEE transactions on medical imaging*, *37*(6), 1348-1357.

Zhao, Y., Li, Y., Dong, X., & Yang, B. (2018). Low-frequency noise suppression method based on improved DnCNN in desert seismic data. *IEEE Geoscience and Remote Sensing Letters*, *16*(5), 811-815.